\documentclass[10pt]{ifcolog-c}

\usepackage{amsmath}
\usepackage{amssymb}
\usepackage{xspace}
\usepackage{color}
\usepackage{graphicx}
\usepackage{enumitem}
\usepackage{todonotes}
\usepackage{booktabs}

\newtheorem{definition}{Definition}
\newtheorem{proposition}{Proposition}

\newtheorem{example}{Example}

\newcommand{\help}[2]{{ [{#1}\! :\! {#2}]}}
\newcommand{\limp}{\rightarrow}
\newcommand{\leqv}{\leftrightarrow}
\newcommand{\defeq}{\stackrel{\text{def}}{=}}
\newcommand{\tuple}[1]{{\langle #1 \rangle}}
\newcommand{\hrel}{{\Rightarrow}}

\newcommand{\Agt}{\textsf{Agt}}
\newcommand{\Atm} {\textsf{Atm} } 

\newcommand{\defined}[1]{{#1}}
\newcommand{\transfun}[1]{\tau^\mathcal{C}(#1)} 
\newcommand{\nepow}[1]{Pow^+(#1)} 
\newcommand{\ndpow}[1]{Pow^+_-(#1)}
\newcommand{\pow}[1]{Pow(#1)} 

\title{A Simple Logic of Cohesive Group Agency\thanks{Originally published in Guillaume Aucher, Jérôme Lang, Tiago de Lima and Emiliano Lorini (eds) \emph{Festschrift for Andreas Herzig on the Occasion of his 65th Birthday: Essays in Honor of Andi}, Tributes, volume 55, College Publications, 2025.}}
\titlerunning{A Simple Logic of Cohesive Group Agency}
\titlethanks{}%Thanks to \ldots}
\addauthor{Nicolas Troquard\thanks{I am grateful to a reviewer for a very helpful report on this contribution. This work was supported by the MUR (Italy) Department of Excellence 2023--2027.}}{Gran Sasso Science Institute, L'Aquila (Italy)}
\authorrunning{Nicolas Troquard}
\nopagenumber

\begin{document}

\maketitle

\begin{abstract}
We propose a structure to represent the social fabric of a
group. We call it the `cohesion network' of the group. It can be seen
as a graph whose vertices are strict subgroups and whose edges
indicate a prescribed `pro-social behaviour' from one subgroup towards
another. In social psychology, pro-social behaviours are building
blocks of full-blown cooperation, which we assimilate here with `group
cohesiveness'. % Constraints on social fabrics define `classes' of cohesion networks.
We then define a formal framework to study cohesive group
agency. To do so, we simply instantiate pro-social behaviour with the
more specific relation of `successful assistance' between
acting entities in a group. The relations of assistance within
a group at the moment of agency constitute the social fabric of
the cohesive group agency.
We build our logical theory upon the logic of agency
``bringing-it-about''. We obtain a
family of logics of cohesive group agency, one for every class of
cohesion networks. % We show that reasoning with the logics is decidable for all classes.
\end{abstract}

%Keywords are your own choice of terms you would like the paper to be indexed by.

% \keywords{group cohesiveness, logic, group agency}

%%\tableofcontents
\section{Introduction}

Group cohesiveness is one of the chief phenomena discussed in social
psychology. Amusingly, the Encyclopedia of Social Psychology has two
entries! One under ``Group cohesiveness''~\cite{eisenberg07}, one
under ``Cohesiveness, Group''~\cite{greifeneder07}. The definitions
proposed there are:
\begin{quote} ``Group cohesiveness (or
  cohesion) is a social process that characterizes groups whose
  members interact with each other and refers to the forces that push
  group members closer together.''~\cite{eisenberg07}
\end{quote}
\begin{quote}
``Cohesiveness refers to the degree of unity or `we-ness' in a
  group. More formally, cohesiveness denotes the strength of all ties
  that link individuals to a group. These ties can be social or task
  oriented in nature. Specifically, a group that is tied together by
  mutual friendship, caring, or personal liking is displaying
  \emph{social cohesiveness}. A group that is tied together by shared
  goals or responsibilities is displaying \emph{task
    cohesiveness}.''~\cite{greifeneder07}
\end{quote}
We will capitalize especially in those forces that tie a group together.
% The ties are not simply between an invidual and the whole group.
Our focus will be on the many ties that contribute to a certain
we-ness---\emph{pro-social behaviours}---that exist within a group and
specifically, between subgroups. The sum of these ties will form the
\emph{social fabric} of the group.
Think of the archetypical group action of lifting a heavy table. The
ties of the group come from each individual helping the rest of the
group to lift the table.

This differs significantly from existing work in social
philosophy and in AI~(e.g.,
\cite{cohen-teamwork91,tuomela00,tuomela07,conte02,dunin10teamwork}). They
address task-related cohesiveness by analysing some combinations of
the powers of agents at doing some sub-task, and of mental attitudes
that agents have towards sub-tasks and other agents. For instance,
group cohesiveness might be manifested when there is a decomposition
of the collective goal in sub-tasks, and appropriately, the members of
the group intend to perform the sub-tasks and they trust each other to
do so.

\medskip

We introduce novel structures that
we coin \emph{cohesion networks}. They are an abstract representation
of the social fabric of a cohesive group realising a collective
goal. 
Our main modelling assumption is that the social fabric of a group $G$ is a
directed graph whose vertices are sub-groups of $G$, and edges
represent pro-social behaviours of a sub-group towards another. 
\begin{example}\label{ex:intro-piano}
Take the action of a group of three agents $1$, $2$ and $3$ lifting a
piano together. We have that $\{1\}$ has a pro-social behaviour
towards $\{2,3\}$, $\{2\}$ has a pro-social behaviour towards
$\{1,3\}$, and $\{3\}$ has a pro-social behaviour towards
$\{1,2\}$. This can be depicted by the following social fabric for
group
$\{1,2,3\}$. \begin{center}
    \begin{tabular}{lcr}
      $\{1\}$ & $\longrightarrow$ & $\{2,3\}$\\ 
      $\{2\}$ & $\longrightarrow$ & $\{1,3\}$\\ 
      $\{3\}$ & $\longrightarrow$ & $\{1,2\}$\\ 
    \end{tabular}
\end{center}
\end{example}

The most simple social fabric exists in groups of two agents.
\begin{example}\label{ex:intro-peanuts}
There is a series of comic strips in Charles Schulz's `The Peanuts' that
always shows ``Lucy van Pelt's interaction with Charlie Brown in the
kick-off practice, in which Lucy is supposed to act as a kick-off tee
and hold the ball steady, while Charlie's part is to run up and kick
the ball.''~(\cite{schmid13})
Supposedly then, the social fabric in this interaction is simply
constituted of a pro-social behaviour from Lucy towards Charlie.
\end{example}

\paragraph{A logic of cohesive group agency}

The nature of individual agency is widely debated in philosophy, or
psychology, where free will, or cognitive dispositions are limiting
cases to fully understanding it. Nonetheless, in many practical cases,
establishing the responsibility of individuals can be judged
uncontroversial. A lone bank robber is caught red-handed; A carpenter
builds a table from scratch in his workshop.

On the other hand, group responsibility~(\cite{BelnapP93,royakkers00})
which is pervasive in legal AI, is often hard to establish. It is
fundamental when one needs, for instance, to establish
responsibilities upon which depend blame and reward.  None of the
gangsters might be deemed individually responsible of the action of
their gang. Yet, the responsibility of the gang, or a part of it, or
several parts of it, could be established.  None of the partners in a
space program might be deemed individually responsible, yet the
responsibility of a consortium could be established, or part of it, or
parts of it. Who in the gang gets to be blamed, and who in the
consortium gets to be rewarded? 

\medskip

If one thinks of responsibility as the fact of bringing about that
some state of affairs is realised, Anselmian's logics of action~(e.g.,
\cite{Prn77actionsocial,elgesem97agency,belnap01facing}) are an
off-the-shelf formal tool to represent responsibility.
Two sub-families exist: the logics of ``bringing-it-about'' and the
logics of ``seeing-to-it-that''. Herzig et al.'s \cite{Herzig2018} contains a succinct introduction to both.

As the examples of the gang and the consortium may hint,
one difficulty lies in the fact that the responsibility of a group is
not indication enough to attribute responsibility to a sub-group, nor
is it to attribute responsibility to a super-group. Yet, all solutions
to group agency in the Anselmian's logics of action go one extreme or
the other. In some, when a group brings about something, then all
super-groups bring it about. This is the case of Kanger \& Kanger's
logic in \cite{kangerkanger66}, and in all but one logic of
\emph{seeing-to-it-that} in \cite{belnap01facing}). In the others,
when a group brings about something, then no strict super-group brings
it about. This is the case in Belnap et
al.~(\cite[Ch~12]{belnap01facing}) and Carmo's~(\cite{carmo10}) logics
of strict joint agency. In~\cite{Norman:2010:LD:1655415.1655567},
interpretations of the agency modality yield either that a coalition
do $\phi$ only if all its members do $\phi$, or that a coalition do
$\phi$ only if one of its members do $\phi$.
None of these extreme stances fit with a commonsense notion of
group responsibility. 
Here, we achieve this by complementing an Anselmian logic of action
with the semantic resource provided by cohesion networks. Hence, we
will deal with group responsibility as \emph{cohesive group agency}.

\section{Social fabrics as cohesion networks}
\label{sec:socialfabric}

In this section, we introduce \emph{cohesion networks}. They are a
formal tool intended to represent the social fabric of a group. A
social fabric for a group is roughly a net of pro-social behaviours
within the group.

The study of pro-social behaviours in prominent in social
psychology. (E.g.,
\cite{darley68bystander,piliavin81,batson98,dovidio06})
\begin{definition}[pro-social behaviour] \label{def:psb}
Informally, a \emph{pro-social behaviour} is
``the broad range of actions intended to benefit one or more people
other than oneself---behaviours such as helping, comforting, sharing
and co-operation.''~\cite[p.~282]{batson98}
\end{definition}
We purposefully maintain the definition of a pro-social behaviour
informal in this section. But for now, we want to suggest that an
underspecified notion of pro-social behaviour is enough to make sense
of a general social fabric. This is only a meta-linguistic object. In
this section, when we say that a pro-social behaviour is
\emph{realised}, this is a fact of the world that needs not to be
interpreted in a more formal way. It does not mean anything more than
the fact that at this moment and in this world, some acting entity
shows a pro-social behaviour towards another. 
\begin{definition}[benefactor / beneficiary]
When a \defined{pro-social\linebreak behaviour} from $C_1$ towards $C_2$ is
realised, we say that $C_1$ is the \emph{benefactor} group and $C_2$
is a \emph{beneficiary} group.
\end{definition}

The intended meaning of pro-social behaviour is different from the
meaning given in~\cite[p.~20]{dovidio06}, where it is ``a broad
category of actions that are `defined by society as generally
beneficial to other people and to the ongoing political
system'~\cite[p.~4]{piliavin81}.'' The pro-social behaviour here is
goal-directed; It is directed towards the goal of some beneficiary.  We
can say that a group of gangsters, who is unlikely to be beneficial to
the ongoing political system, is acting cohesively, and thus the
gangsters demonstrate a pro-social behaviour \emph{within the gang}.

Later in Section~\ref{sec:agency} we will formally instantiate the
notion of pro-social behaviour with a specific kind of helping
behaviour and it will be interpreted on unambiguous formal models.

\medskip

Throughout the paper, we will assume a finite set of individual agents
denoted by $\Agt$. A group of agents could be simply defined as an
arbitrary set of agents in $\Agt$. However, we prefer here refusing
the right to the empty set to be a group of agents. A \emph{group} of
agents will then be a non empty element of the powerset of $\Agt$,
noted $\nepow{\Agt}$. To make the formulas lighter, we will
occasionally write simply $i$ instead of $\{i\}$, where $i \in \Agt$.

Delving into a notion of group cohesiveness, another particular case
must be accounted for. A singleton group, that is $\{i\}$ for some
individual agent $i \in \Agt$, is a group of agents, although a
degenerate one. But we will not want to assign to it a group
cohesiveness proper. We expect from an individual to act with some
sense of \emph{coherence} (which is beyond the scope of this paper)
but not \emph{cohesively} `within itself'. We note $\ndpow{\Agt}$ the
set of non-degenerate groups $G \in (\nepow{\Agt}\setminus \{\{i\}
\mid i \in \Agt\})$.

\medskip

\begin{definition}[cohesion networks]\label{def:cohnet}
A \emph{cohesion network} for\linebreak $G \in \ndpow{\Agt}$ is a tuple $\tuple{\Gamma, \hrel}$ such that:
\begin{enumerate}[itemsep=0pt]
\item \label{cons:graph1} $\Gamma \subseteq \pow{G}$;
\item \label{cons:graph2} $\hrel \subseteq \Gamma \times \Gamma$;
\item \label{cons:noAll} $G \not \in \Gamma$;
\item \label{cons:noNone} $\emptyset \not \in \Gamma$.
\item \label{cons:allin} $G \subseteq \{C_1, C_2 \mid (C_1,C_2) \in \hrel\}$.
\end{enumerate}
\end{definition}
We note $\mathcal{C}_0(G)$ the set of all admissible cohesion networks for $G$.

Based upon our newly defined cohesion networks and an arbitrary
understanding of a \defined{pro-social behaviour} fitting
Definition~\ref{def:psb}, we can provide the definition of the notion
of \emph{cohesiveness} addressed in this paper.

In the interest of clarity, we introduce some additional terminology
before we explain the constraints of Definition~\ref{def:cohnet}.
\begin{definition}[witness, cohesiveness, reliance, wrt.\ $\mathcal{C}_0$]\label{def:witcohe}
We say that an admissible cohesion network $\tuple{\Gamma, \hrel} \in
\mathcal{C}_0(G)$ is a \emph{witness for cohesiveness}
(wrt.\ $\mathcal{C}_0$) of $G$ when for all $(C_1, C_2) \in \hrel$ there
is a realised \defined{pro-social behaviour} from $C_1$ towards $C_2$.
A group of agents $G$ is said to be \defined{cohesive} (wrt.\
$\mathcal{C}_0$) if there is a \defined{cohesion network} in
$\tuple{\Gamma, \hrel} \in \mathcal{C}_0(G)$ which is
a \defined{witness for cohesiveness} of $G$.

If $(C_1, C_2) \in \hrel$ for some $\tuple{\Gamma, \hrel} \in
\mathcal{C}_0(G)$, then we say that to be \defined{cohesive} (wrt.\ $\mathcal{C}_0$), $G$
\emph{may rely} on a \defined{pro-social behaviour} from
$C_1$ towards $C_2$.
If for all $\tuple{\Gamma, \hrel} \in \mathcal{C}_0(G)$ there is 
$(C_1, C_2) \in \hrel$ such that $i \in C_1 \cup C_2$ then we say that
to be \defined{cohesive} (wrt.\ $\mathcal{C}_0$), $G$ \emph{must rely} on the agent $i$.
\end{definition}
An admissible cohesion network for a group $G$ is meant to capture an
admissible social fabric that is sufficient to deem the group $G$
cohesive, provided that all prescribed \defined{pro-social behaviours} are
realised.

Constraint~\ref{cons:graph1} and Constraint~\ref{cons:graph2} imply that a group $G$ \defined{may not rely} on outsider agents to be \defined{cohesive}.
Constraint~\ref{cons:noAll} enforces a (critical for Section~\ref{sec:agency}!) reductionist view of group cohesiveness. It says that
for a group $G$ to be \defined{cohesive}, it \defined{may not rely} on a
\defined{pro-social behaviour} involving $G$.
(Still, it \defined{may rely} on a \defined{pro-social behaviour} involving all the members of $G$.) 
Similarly, Constraint~\ref{cons:noNone} says that
it \defined{may not rely} on a \defined{pro-social behaviour} involving the empty coalition either.
Constraint~\ref{cons:allin} says that a group $G$ to be
\defined{cohesive}, it \defined{must rely} on all its members.

\begin{example}\label{ex:C0-2ag}
The smaller groups to have a social fabric are groups with two
members. There are exactly three cohesion networks that are admissible
wrt.\ $\mathcal{C}_0$ for each of such group. Let $\Gamma = \{\{1\},
\{2\}\}$. 
We have $\mathcal{C}_0(\{1,2\}) = \{ \tuple{\Gamma,
  \{(\{1\},\{2\})\}}, \tuple{\Gamma, \{(\{2\},\{1\})\}},
%\linebreak \tuple{\Gamma, \{(\{1\},\{2\}), (\{2\},\{1\})\}} \}$.
\langle \Gamma, \{(\{1\},\{2\}),\linebreak (\{2\},\{1\})\} \rangle \}$.
So a group $\{1,2\}$ can be \defined{cohesive} when one of the following is
realised: $1$ has a \defined{pro-social behaviour} towards~$2$,
when~$2$ has a \defined{pro-social behaviour} towards~$1$, or when
both~$1$ and~$2$ has a \defined{pro-social behaviour} towards the
other.

A group of three agents has already many ways of being \defined{cohesive}. For instance $\tuple{\Gamma, \{(\{1\}, \{2\}), (\{1,2\},\{3\})\}}$ is admissible wrt.\ $\mathcal{C}_0(\{1,2,3\})$, and so is
$\tuple{\Gamma, \{(\{1\}, \{2,3\}), (\{2\}, \{1,3\}), (\{3\}, \{1,2\})\}}$ from Example~\ref{ex:intro-piano}.
\end{example}

\medskip

In practice, a system designer would have to design a class of
cohesion networks that reflects the notion of social fabric that is
relevant for the application at hand. Specific classes of cohesion
networks can be defined by constraining $\mathcal{C}_0$ further.
Obviously, $\mathcal{C}_0$ is a class of cohesion networks.
\begin{definition}[class of cohesion network]
A \emph{class of cohesion networks} is an object $\mathcal{C}$ such
that for every group $G \subseteq \Agt$ we have $\mathcal{C}(G) \subseteq \mathcal{C}_0(G)$.
\end{definition}

\section{Individual agency and successful assistance}
\label{sec:background}

Logics of agency are the logics of modalities $E_x$ for where $x$ is
an acting entity, and $E_x\phi$ reads ``$x$ brings about $\phi$'', or
``$x$ sees to it that $\phi$''. This tradition in logics of action
comes from the observation that action can be explained by what it
brings about. See \cite{belnap88theoria,belnap01facing}.
Here, we will specifically work with a logic of
bringing-it-about (BIAT).  It has been studied over several decades in
philosophy of action, AI law, and in multi-agent systems
(\cite{kangerkanger66}, \cite{Prn77actionsocial}, \cite{lindahl77},
\cite{Elgesem93phd}, \cite{CarmoDEON96}, \cite{santos97},
\cite{elgesem97agency}, \cite{royakkers00}, \cite{sartor04}, \cite{carmo10},
\cite{Tr12aamas}, \cite{sartor12}, \cite{DBLP:journals/aamas/Troquard14}, \cite{DBLP:conf/ecai/PorelloT14}).  The philosophy that grounds the
logic was carefully discussed by Elgesem
in~\cite{Elgesem93phd}.
Borrowing from~\cite{santos97}, we will also integrate one modality
$A_x$ (originally noted $H_x$) for every acting entity $x$, and
$A_x\phi$ reads ``$x$ tries to bring about $\phi$''.  Lorini \&
Herzig~(\cite{lorini08synthese}) observe that $A_x\phi$ reflects
Schroeder's conceptualisation of trying~(\cite{schroeder01}).
That is, $A_x\phi$ is merely the judgment from the point of view of an external observer 
that $x$ tries to exercise his control towards $\phi$, but
$x$ may fail to exercise this control proper.

\medskip

We assume a finite set of agents $\Agt$ and an enumerable set of
atomic propositions $\Atm$. The language of BIAT extends the language
of propositional logic over $\Atm$, with one operator $E_i$ and one
operator $A_i$ for every agent $i \in \Agt$. 

\begin{center}
\begin{tabular}{ll}
(prop) & $\vdash_{BIAT} \phi$ \hfill, when $\phi$ is a classical tautology\\
(notaut) & $\vdash_{BIAT} \lnot E_i\top$\\
%(aggreg) & $\vdash E_x \phi \land E_x \psi \limp E_x (\phi \land \psi)$
(success) & $\vdash_{BIAT} E_i \phi \limp \phi$\\
%(attempt) & $\vdash_{BIAT} E_i \phi \limp A_i\phi$\\\\
(ree) & if $\vdash_{BIAT} \phi \leqv \psi$ then $\vdash_{BIAT} E_i \phi \leqv E_i \psi$\\
(rea) & if $\vdash_{BIAT} \phi \leqv \psi$ then $\vdash_{BIAT} A_i \phi \leqv A_i \psi$
\end{tabular}
\end{center}
BIAT extends propositional classical logic (prop). An acting entity
never exercises control towards a tautology (notaut).
Agency is an achievement, that is, the culmination of
a successful action (success).
Agency and attempts are closed under provably equivalent formulas (ree) and
(rea). 
The satisfiability problem in BIAT is decidable~\cite{vardi89,DBLP:journals/aamas/Troquard14}.

\medskip

We have been concerned about pro-social behaviour in the first part of
this paper. Here, we  define an event of successful
assistance---a particular kind of pro-social behaviour.

Helping behaviour is defined in social psychology as ``an action that has the consequence of providing some benefit to or improving the well-being of another person''~\cite[p.~22]{piliavin81}.
Tuomela~\cite[p.~86]{tuomela00} explains how help events are found as constituting parts of cooperative actions.
Specifically, our events of assistance will be of the nature of
contributing to or participating in a resulting state of affairs,
possibly only by counteracting negative interference. 
We define a new modality of agency: $\help{i}{j}\phi$. It is intended
to read ``the agent $i$ \emph{successfully assists} the agent $j$
to achieve $\phi$''. 
\[\help{i}{j}\phi \defeq E_{i} (A_j \phi \limp \phi) \land A_j\phi\]
Literally, agent~$i$ brings about that if agent~$j$ tries to achieve $\phi$ then $\phi$ holds, and agent~$j$ does try to achieve $\phi$. This is studied in great details in \cite{Bottazzi2015}.

It is a \emph{successful} assistance because we have the following expected
property by applying (success) and (prop):
%\begin{proposition}
$\vdash_{BIAT} \help{i}{j}\phi \limp \phi$.
%\end{proposition}
%
It is an event of \emph{assistance}, for three reasons. First, there
is an \emph{assistee}. It is a goal of $j$ to bring about $\phi$ as $j$
does try. Second, there is an \emph{assistant}. $i$'s guidance is
reactive to $j$'s goodwill in the action. Here, the goal of $i$ is
that $\phi$ holds if $j$ tries to bring about $\phi$. Third, it is
compelling to a formalisation of assistance that $\help{i}{j}\phi
\land \lnot E_{i}\phi \land \lnot E_{j}\phi$ is a consistent formula.
That is, it is possible that $i$ successfully assists $j$ to bring
about $\phi$, and still, neither $i$ nor $j$ brings about $\phi$.
Hence, the success of the event of assistance described by
$\help{i}{j}\phi$ comes from some cohesion between $i$ and $j$.

\section{The logic of cohesive group agency}
\label{sec:agency}

In this section, we define one logic of cohesive group agency for every
class of cohesion networks. We investigate some properties of the logics.

As before, we assume a finite set of individual agents $\Agt$ and a
finite set of atomic propositions $\Atm$. The language $L$ is defined
by the following grammar:
\[
\begin{array}{lccccccccccccc}
\phi & ::= & p & \mid &\lnot \phi & \mid & \phi \land \phi & \mid\\
& & E_G\phi &
\mid & A_G\phi & \mid & \help{C_1}{C_2}\phi\\
\end{array}
\]
where $p \in \Atm$, and $G, C_1, C_2 \in \nepow{\Agt}$.

As previously, $E_{G}\phi$ means that ``$G$ brings about that
$\phi$''. For $G \in \ndpow{\Agt}$, one may read ``$G$ 
cohesively brings about that $\phi$''. This section describes the
formal machinery that justifies this reading.

\medskip

For there to be full-blown group agency, it is a platitude to say
there must be full-blown cooperation. ``Full blown cooperation is
based on a shared collective goal and requires acting
together''~\cite[p.~372]{tuomela00}. Echoing the case of individual
agency reported above, Miller~(\cite{miller01social}) acknowledges
that shared agency is directed towards a goal, but argues that shared
agency does not require shared intention. In
consequence, group agency is oriented towards a collective goal, but
the group does not have to be collectively aware of this
goal. 

What then can support a claim of group agency for a state of affairs?
We propose to base the interpretation of group agency on the social
fabric of a group. Cohesion networks are general tools to represent a
social fabric of a group. In the remaining of this paper, we will
apply them specifically to cohesive group agency.

For every class of cohesion networks $\mathcal{C}$, our aim here is to
formalise ``$G$ is cohesively agentive for $\phi$'' to correspond to
the fact that there is a cohesion network $\tuple{\Gamma, \hrel}
\in\mathcal{C}(G)$ such that if $(C_1, C_2) \in \hrel$ then there is a
\defined{successful assistance} from $C_1$ towards $C_2$ for
obtaining $\phi$.

\subsection{Cohesively bringing about}

We adopt a reductionist view of group agency. That is, we intend to
explain what a group brings about in terms of the agentive attitudes
of its subgroups. To do so, we are going to define a function
$\transfun{.} : L \longrightarrow L$ that transforms a formula of
$\phi \in L$ into a formula of $\transfun{\phi} \in L$ containing no
occurrence of $E_G$ with non-degenerate group $G$. We detail now how
$\transfun{.}$ defines the three modalities of the language by mutual
induction.

The definition of $\help{C_1}{C_2}\phi$ mirrors the definition of
successful help between two individuals in BIAT.
\begin{equation}\label{eq:help}
\transfun{\help{C_1}{C_2}\phi} = \transfun{E_{C_1} (A_{C_2}\phi \limp \phi) \land A_{C_2}\phi}
\end{equation}
Cohesive group agency for a state of affairs $\phi$ is the special
case of group cohesiveness defined in Definition~\ref{def:witcohe},
where a \defined{pro-social behaviour} from $C_1$ towards $C_2$ is
exactly the event of $C_1$ successfully helping $C_2$ to bring about
$\phi$. As we can obtain different accounts of group cohesiveness
depending on the class of cohesion network we use, we will also have
one notion of cohesive group agency for each of them. Then, given a
class $\mathcal{C}$, we say that $G$ cohesively brings about $\phi$ if
there is a cohesion network $\tuple{\Gamma, \hrel} \in \mathcal{C}(G)$
such that for all $(C_1, C_2) \in \hrel$, $C_1$ successfully helps
$C_2$ to achieve $\phi$. In formula, we define:
\begin{equation}\label{eq:defcohag}
\transfun{E_G \phi} = \bigvee_{\tuple{\Gamma,\hrel} \in \mathcal{C}(G)} \bigwedge_{C_1 \hrel C_2} \transfun{\help{C_1}{C_2}\phi}, \text{ when } G \in \ndpow{\Agt}
\end{equation}
Additionally, we consider that a group $G$ attempts to bring about $\phi$ iff it is the attempt of all singleton coalitions in $G$.\footnote{This is arguably an over-simplying view on group attempts. However, the approach presented here is amenable to any reductionist definition of group attempt.}
\begin{equation}\label{eq:Gattempt}
\transfun{A_G\phi} = \bigwedge_{i \in G} A_{\{i\}}\transfun{\phi}, ~~ \text{ when } G \in \ndpow{\Agt}
\end{equation}

\subsection{Axiomatisation}
Given a class of cohesion networks $\mathcal{C}$, the proof theory
$\vdash_\mathcal{C}$ is summarised in Table~\ref{tab:axioms}.
\begin{table}[h]
\begin{center}

\framebox{\small
\begin{tabular}{ll}
(prop) & $\vdash_\mathcal{C} \phi$ \hfill, when $\phi$ is a classical tautology\\
(notaut) & $\vdash_\mathcal{C} \lnot E_{\{i\}}\top$\\
(success) & $\vdash_\mathcal{C} E_{\{i\}} \phi \limp \phi$\\
%% (emptye) & $\vdash_\mathcal{C} \lnot E_\emptyset \phi$\\
%% (emptya) & $\vdash_\mathcal{C} A_\emptyset \phi$\\
%(attempt) & $\vdash_\mathcal{C} E_G \phi \limp A_G\phi$\\
(help) & $\vdash_\mathcal{C} \help{C_1}{C_2}\phi \leqv E_{C_1} (A_{C_2}\phi \limp \phi) \land A_{C_2}\phi$\\
(cohagen) & $\vdash_\mathcal{C} E_G \phi \leqv \bigvee_{\tuple{\Gamma,\hrel}
  \in \mathcal{C}(G)} \bigwedge_{C_1 \hrel C_2} \help{C_1}{C_2}\phi$ \hfill , $G \in \ndpow{\Agt}$\\
(attind) & $\vdash_\mathcal{C} A_G \phi \leqv \bigwedge_{i \in G} A_{\{i\}}\phi$ \hfill , $G \in \ndpow{\Agt}$\\
(ree) & if $\vdash_\mathcal{C} \phi \leqv \psi$ then $\vdash_\mathcal{C}
  E_{\{i\}} \phi \leqv E_{\{i\}} \psi$\\
(rea) & if $\vdash_\mathcal{C} \phi \leqv \psi$ then $\vdash_\mathcal{C}
  A_{\{i\}} \phi \leqv A_{\{i\}} \psi$
\end{tabular}
}
\end{center}
\caption{$\vdash_\mathcal{C}$\label{tab:axioms}}
\end{table}
Axioms (cohagen), (help), and (atting) merely mimic, respectively,
Equation~\ref{eq:defcohag}, Equation~\ref{eq:help}, and
Equation~\ref{eq:Gattempt}. Principles~(notaut), (success), (ree), and
(rea) ensure that $E_{\{i\}}$ and $A_{\{i\}}$ behave like in BIAT.
These properties generalise to $E_G$ and $A_G$.

The logic of cohesive group agency is decidable for every class of cohesion network.
\begin{proposition}\label{prop:dec}
Let a formula $\phi \in L$. For any class of cohesion networks
$\mathcal{C}$, there is an algorithm to decide whether
$\vdash_\mathcal{C}\phi$.
\end{proposition}
Indeed, Constraint~\ref{cons:noAll} of Definition~\ref{def:cohnet} ensures that
every formula can be reduced to a formula with only singleton coalitions.
A formula with only singleton coalitions is equivalent to a BIAT formula, where every coalition $\{i\}$ is replaced with agent~$i$. The result then follows from the decidability of BIAT~\cite{DBLP:journals/aamas/Troquard14}.

\subsection{Example I: Piano}
\label{sec:ex-piano-devel}

Consider now a continuation of Example~\ref{ex:intro-piano}.  The only
admissible cohesion network for the group $\{1,2,3\}$ is the one where
each individual has a pro-social behaviour towards the group formed by
the two others. We can formalise the statement that $\{1,2,3\}$ bring
about that the piano is lifted. Suppose $p$ stands for ``the piano is
lifted''. Recursively applying (help), (cohagen) and (attind) we obtain:
\begin{itemize}[itemsep=0pt]
\item $E_{\{1,2,3\}}p \leqv
  \help{\{1\}}{\{2,3\}}p \land
  \help{\{2\}}{\{1,3\}}p \land
  \help{\{3\}}{\{1,2\}}p$
\item $E_{\{1,2,3\}}p \leqv E_1(A_2 p \land A_3 p \limp p)
\land E_2(A_1 p \land A_3 p \limp p) \land 
E_3(A_1 p \land A_2 p \limp p) \land A_1 p \land A_2 p \land A_3 p$
\end{itemize}
That is, the group ${\{1,2,3\}}$ brings about that the piano is lifted
iff each individual tries to bring about that the piano is lifted, and
each individual brings about that if both other individuals try to
bring about that the piano is lifted, then the piano is lifted.

\subsection{Example II: Peanuts}
Let us go back to the situation of the football gag sketched in
Example~\ref{ex:intro-peanuts}.  Invariably in the cartoons, Charlie
would run towards the ball and fail to hit the ball. What must happen
for the failure of the cooperative action between Charlie and Lucy?
This is captured simply by the formula $\lnot E_{\{Charlie,Lucy\}} k$,
where $k$ stands for ``the ball is kicked by Charlie''.

Since we are looking for reasons for failure, it may be better to not
concentrate on a specific class of cohesion network. So we assess the
situation with respect to the most general class of cohesion networks.
$$\lnot E_{\{Charlie,Lucy\}} k \leqv
\lnot\bigvee_{\tuple{\Gamma,\hrel} \in
  \mathcal{C}_0(\{Charlie,Lucy\})} \bigwedge_{C_1 \hrel C_2}
\help{C_1}{C_2}k$$ There are three possible cohesion networks for
$\{Charlie,Lucy\}$ wrt.\ $\mathcal{C}_0$. Hence, $\{Charlie,Lucy\}$
brings about that the ball is kicked by Charlie iff one the following
is the case~(cf. Example~\ref{ex:C0-2ag}):
\begin{enumerate}[itemsep=0pt]
\item Charlie successfully assists Lucy to bring about $k$
\item Lucy successfully assists Charlie to bring about $k$
\item $1$ and $2$
\end{enumerate}
Instead, $\lnot E_{\{Charlie,Lucy\}} k$ holds iff none of the above
holds. Hence, the failure of the cooperative action is due to the fact
that Charlie does not help Lucy to bring about $k$, \emph{and} Lucy does not
help Charlie to bring about $k$.

The dialogue between Lucy and Charlie suggests that if the cooperation
were to be successful, that is if ${\{Charlie,Lucy\}}$ were to be
cohesively bringing about $k$, then Lucy would have to successfully
assist Charlie to bring about $k$. Clearly, the strips story hints at
the fact that Lucy really does not help Charlie in the matter:
$\lnot\help{\{Lucy\}}{\{Charlie\}}k$. That is, $\lnot
E_{\{Lucy\}}(A_{\{Charlie\}}k \limp k) \lor \lnot
A_{\{Charlie\}}k$. It seems obvious that Charlie does try to bring
about that he kicks the ball. Schmid~\cite{schmid13} even qualifies it as a confident
trying. So we can turn our attention to the feature that must be
incriminated: Lucy does not bring about that the ball is kicked by
Charlie if he tries. Indeed, most of the time, Lucy pulls the ball
away at the last moment. But she does not always fail to bring about
$A_{Charlie} k \limp k$ maliciously: in the strip of the 16th of
November 1952 for instance, the assistance fails because she holds the
ball ``real tight''. Too tight.

Also, it must be the case that Charlie does not successfully assist
Lucy to bring about $k$: $\lnot\help{\{Charlie\}}{\{Lucy\}}k$. That
is,\linebreak $\lnot E_{\{Charlie\}}(A_{\{Lucy\}}k \limp k) \lor \lnot A_{\{Lucy\}}k$.
Sometimes, the story clearly suggests that the second
disjunct of the previous formula is true: Lucy does not try to bring
about that the ball is kicked by Charlie. 
The fact $E_{\{Charlie\}}(A_{\{Lucy\}}k \limp k)$ might be
true in the story: Charlie brings about that he kicks the ball if Lucy
tries to bring about that the ball is kicked by Charlie. No matter
what, this is not enough to save the situation as long as Lucy does
not try to bring about $k$.

\section*{Andreas}

Andi is one of the main agents in the field of logics for AI, especially logics for agents and multiagent systems. His work evidences his choice with commitment towards revisiting ideas \cite{DBLP:conf/kr/HerzigL04}, and he often sees to it that solid bridges are built with other disciplines, such as philosophy \cite{DBLP:journals/logcom/BroersenHT06,DBLP:journals/jphil/HerzigL11}.
One of the more specific topics that pique his interest is figuring out the formal and computational dynamics of groups and institutions of agents \cite{DBLP:conf/atal/HerzigL02,DBLP:journals/jancl/BroersenHT09,DBLP:journals/synthese/HerzigLL09,DBLP:journals/logcom/LoriniLGH09,DBLP:conf/mabs/GaudouHLS11,DBLP:conf/aiml/HerzigP20}.
Moreover, he always keeps a certain taste for simplicity \cite{DBLP:conf/wollic/Herzig13,DBLP:conf/lori/HerzigLM15,DBLP:conf/ecai/CooperHMMR16a}

Andi's approach to research has influenced the way I conduct my own work more than anyone or anything else. I hope he likes this simple logic, which is a revisitation of the formal aspects of group agency, with some light connections to the social sciences.

%%\def\newblock{}
%%\bibliographystyle{named}
%%\small
\bibliographystyle{plain}
\bibliography{biblio.bib}

\end{document}